\documentclass[12pt]{article}

\newread\testifexists \def\GetIfExists #1
{\immediate\openin\testifexists=#1
    \ifeof\testifexists\immediate\closein\testifexists\else
    \immediate\closein\testifexists\input #1\fi}

\usepackage{amsfonts} \usepackage{amssymb}
\immediate\openin\testifexists=e
    \ifeof\testifexists\immediate\closein\testifexists\else
    \immediate\closein\testifexists\input e\fipsf
\def\Bbb#1{\setbox0=\hbox{$\tt #1$} \copy0\kern-\wd0\kern .1em\copy0}
\def\bbf#1{\setbox0=\hbox{$#1$} \kern-.025em\copy0\kern-\wd0
        \kern.05em\copy0\kern-\wd0 \kern-.025em\raise.0433em\box0}
 \immediate\openin\testifexists=a
    \ifeof\testifexists\immediate\closein\testifexists\else
    \immediate\closein\testifexists\input a\fimssym.def  % for \Bbb A - Z %%

\def\a{\alpha} \def\b{\beta} \def\g{\gamma} \def\G{\Gamma} \def\d{\delta}  \def\e{\varepsilon}
\def\et{\eta} \def\k{\kappa} \def\l{\lambda} \def\L{\Lambda} \def\m{\mu} \def\f{\phi} 
 \def\n{\nu}    \def\s{\sigma} 
      
\def\w{\omega}  

 \def\LL{{\cal L}} \def\OO{{\cal O}}\def\DD{{\cal D}}
\def\pa{\partial} \def\ra{\rightarrow} 
  \def\dd{{\rm d}}
\def\bra{\langle} \def\ket{\rangle}

\def\fract#1#2{{\textstyle{#1\over#2}}} \def\ffract#1#2{\raise .3
em\hbox{$\scriptstyle#1$}\kern-.25em/
                \kern-.2em\lower .2 em \hbox{$\scriptstyle#2$}}
\def\fractje#1#2{{\scriptstyle{#1\over#2}}} \def\half{\fract12}
\def\quart{\fract14}  
\def\part#1#2{{\partial#1\over\partial#2}}

\def\iz{\quad = \quad}\def\iss{\ =\ }

\newcommand{\tl}[1]{\tilde{#1}} 
 \newcommand{\be}{\begin{eqnarray}}
\newcommand{\ee}{\end{eqnarray}} \newcommand{\eqn}[1]{(\ref{#1})}
\newcommand{\nn}{\nonumber\\} \newcommand{\nm}{\nonumber}

\newcommand{\bi}[1]{\begin{itemize}\item[#1]}
 \newcommand{\ei}{\end{itemize}}

\newcommand{\fn}{\footnote}
\newcommand{\crlb}[1]{\label{#1}\\[2pt]}
\newcommand{\eela}[1]{\quad\hbox{\scriptsize{#1}}\label{#1}\end{eqnarray}}
\newcommand{\eelb}[1]{\label{#1}\end{eqnarray}}

\newcommand{\newsecb}[2]{\section{#1}\label{#2}\setcounter{equation}{0}}

\newcommand{\nolabels}{ \def\eel{\eelb} \def\crl{\crlb} \def\newsecl{\newsecb} }

\def\publishversion{\nolabels\setlength{\textheight}{9in}\setlength{\oddsidemargin}{0in}
    \setlength{\textwidth}{6.3in}\setlength{\topmargin}{-0.1in}}

\publishversion 
\begin{document} \begin{titlepage}

\title{\normalsize \hfill ITP-UU-07/44 \\ \hfill SPIN-07/32 \\ \vskip 5mm \Large\bf Unitarity in
the Brout-Englert-Higgs Mechanism for Gravity}

\author{Gerard 't~Hooft}
\date{\normalsize Institute for Theoretical Physics \\
Utrecht University \\ and \medskip \\ Spinoza Institute \\ Postbox
80.195 \\ 3508 TD Utrecht, the Netherlands \smallskip \\ e-mail: \tt
g.thooft@phys.uu.nl \\ internet: \tt
http://www.phys.uu.nl/\~{}thooft/}

\maketitle

 \noindent {\large\bf Abstract}\medskip

Just like the vector bosons in Abelian and non-Abelian gauge theories, gravitons can attain mass by
spontaneous local symmetry breaking. The question is whether this can happen in a Lorentz-invariant way. We
consider the use of four scalar fields that break coordinate reparam\-etriz\-ation invariance, by playing the
role of preferred flat coordinates $x$, $y$, $z$, and $t$. In the unbroken representation, the theory has a
(negative) cosmological constant, which is tuned to zero by the scalars in the broken phase. Massive spin 2
bosons and a single massive scalar survive. The theory is not renormalizable, so at best it can be viewed as
an effective field theory for massive spin 2 particles. One may think of applications in cosmology, but a
more tantalizing idea is to apply it to string theory approaches to QCD: if the gluon sector is to be
described by a compactified 26 or 10 dimensional bosonic string theory, then the ideas considered here could
be used to describe the mechanism that removes a massless or tachyonic scalar and provides mass to the spin 2
glueball states. The delicate problem of removing indefinite metric and/or negative energy states is
addressed. The scalar particle has negative metric, so that unitarity demands that only states with an even
number of them are allowed. Various ways are considered to adapt the matter section of the theory such that
matter only couples to positive metric states, and we succeed in suppressing the main contributions to
unitarity-violating amplitudes, but the exact restoration of unitarity in the spinless sector will continue
to be a delicate issue in theories of this sort.

%\vfill \flushleft{\today}

\end{titlepage}

\eject

%%%%%%%%%%%%%%%%%%%%%%%%%%%%%%%%%%%%%%%%%%%%%%%%%%%%
\newsecl{Introduction: Motivations for this study}{intro.sec}
%%%%%%%%%%%%%%%%%%%%%%%%%%%%%%%%%%%%%%%%%%%%%%%%%%%%
There are two, almost disconnected, areas in theoretical physics where it may be worthwhile to consider the
spontaneous breakdown of general coordinate reparametrization invariance. One is the study of cosmological
models. Massless scalar fields, \emph{only} interacting with gravity, could exist in principle. If so, they
can give mass to the graviton, thus neutralizing the gravitational force at large distances. There have been
numerous speculations on such effects in the literature, but they usually suffer from stability and
fine-tuning problems. A tiny cosmological constant would be neutralized by these scalars, and they would lead
to a universe that has a tendency to remain flat. The cosmological constant problem itself concerns the
occurrence of vastly different scales in physics, and this problem would not be resolved with such ideas, so
skepticism is justified, but further investigation along these lines could be of interest.

A second motivation for our work could be the study of Quantum Chromodynamics (QCD). The forces generated by
the gluon fields take the form of vortex tubes. If we leave out the quarks, these vortices become
unbreakable, and behave like bosonic strings. However, bosonic string theory is beset by anomalies in four
space-time dimensions. There are no such anomalies in QCD. At best, therefore, QCD can be approximated by a
string theory where the strings are disfigured by non minimal self-interactions, even in the \(N\ra\infty\)
limit. Is there no way to describe such self-interacting strings by starting from the powerful string theory
approaches that exist today? Since the gluonic sector has no fermionic states, the first candidate to think
of as a starting point would be a 26 dimensional bosonic string theory\cite{bosstring}, of which 22
dimensions are compactified. However, this theory has tachyonic and massless scalar (spin 0), massless vector
(spin 1) and massless tensor (spin 2) solutions\cite{Siegel93}. None of these are expected to exist in the
pure gluonic sector of QCD.\fn{Note, that what we have in mind is {not} to {replace} QCD by a string theory,
but rather to use string theory techniques to do calculations in the sector of confined gluons and, at a
later stages, confined quarks in conventional QCD.}

As for the massless vector particles, there is an elegant way to remove these: the Brout-Englert-Higgs (BEH)
mech\-an\-ism.\cite{BEH} Due to interactions, the tachyonic scalar (in the symmetric representation) develops
a vacuum expectation value, after which the various states rearrange: some of the scalar fields combine with
the massless vector fields to produce the dynamical spin 1 solutions of a massive spin 1 field, exactly as it
happens in the electro-weak sector of the Standard Model.

However, the string theory also has massless spin 2 modes, which somehow disappear in QCD. Naturally, one
would like to invoke a mechanism analogous to the BEH mechanism. The author has been aware of this
possibility for some time, but only now it was realized how one might proceed to overcome a nasty problem
with indefinite metric and/or negative energy states. Stability of the vacuum requires that all fields
describing elementary particles have only excitation modes with non-negative energies. This requires not only
the absence of tachyonic mass terms, but also non-negative coefficients in all kinetic terms of the
Lagrangian, and of course the existence of an unambiguous inverse of the bilinear terms that can serve as
propagators. If one attempts to quantize a theory with negative coefficients in the kinetic terms, one
encounters quantum states with indefinite norm, thus violating unitarity, an important consistency condition
in quantum field theories\cite{DIAGRAMMAR}. We defer the discussion of this problem to
Sections~\ref{indefinite.sec} and onwards; let us momentarily assume that the problem can be handled. Thus,
we may have achieved an opening towards describing an interacting string theory containing only massive
string modes. One will also have to describe self interactions between strings, so it will not be an easy
solution, but at least we see how, in principle, massless modes can be avoided. Indeed, one also would avoid
supersymmetry this way, so that we avoid the nonexisting fermionic states in the gluonic sector of QCD.

These arguments should not be regarded as opposed to, but rather complimentary to the AdS/CFT approach to
solve QCD using superstring theory\cite{Kiritsis}, where the 3+1 dimensional theory is mapped onto a 5
dimensional AdS theory. There, the massless graviton in 5 dimensions is mapped onto a massive graviton in 4
dimensions. The five-dimensional theory may have to be supersymmetric again\fn{I thank E.~Kiritsis for an
enlightening explanation of these points.}. What we try to do here is develop a conceptual understanding of
what happens in 4 dimensions.

There is in fact another problem standing in the way of a string theoretical approach to QCD. QCD approaches
its perturbative approximation in the far ultraviolet domain, where all gluonic states are far from the mass
shell. It is this domain, where the basic QCD action is precisely defined, and where one would like to
compare QCD field variables with string degrees of freedom. If we can find a match (by some mathematical
transformation), then we have the beginning of a systematic procedure to identify string amplitudes with QCD
amplitudes. Yet string theory only allows identification of its on-mass-shell states. Possibly, the deeper
reason why string theory does not allow a consistent off-shell description is the fact that, in its standard
representation, string theory is invariant under coordinate reparametrizations. So, observable coordinates do
not exist, which may well be the reason why local fields cannot exist --- we would not be able to specify
their coordinates in a meaningful way.

However, when the coordinate reparametrization invariance is spontaneously broken, we \emph{do} have
coordinates. Our four scalar fields will serve as such. Thus, being able to provide for gauge-invariant
coordinates, perhaps ``off shell" amplitudes can now be defined. This could open the door even further
towards a consistent way of treating QCD using string theory, while avoiding the unphysical supersymmetric
theories that have been put forward until now.

Most of our discussion is limited to 4-dimensional space-time, with asymptotically flat boundary conditions.

\newsecl{Perturbative Einstein-Hilbert gravity with massless scalars}{EHgrav.sec}

Let us begin by establishing our notation. For a more elaborate description of perturbative
(quantum) gravity see Ref.~\cite{Erice02}. The pure Einstein-Hilbert action is
 \be S=\int\LL(x)\,\dd^4x\ ;\qquad \LL(x)={\sqrt{g}\over 16\pi G} (R-2\L)+\LL_1\ .\eel{EH}
\(R\) is the Ricci scalar curvature, \(g\) is the determinant of the metric tensor \(g_{\m\n}\) (in
Euclidean notation), \(\LL_1\) is a remainder, to be discussed later. \(\L\) is a possible
cosmological constant. In this section, we shall keep \(\L=0\). We have the usual definitions:
 \be \G_{\a\m\n}=\half(\pa_\m g_{\a\n}+\pa_\n g_{\a\m}-\pa_\a g_{\m\n})\ ;\qquad
\G^\l_{\m\n}=g^{\l\a}\G_{\a\m\n}\ .\\ R^\l_{\ \a\m\n}=\pa_\m\G^\l_{\a\n}-\pa_\n\G^\l_{\a\m}+\G^\l_{\m\s}
\G^\s_{\a\n}-\G^\l_{\n\s}\G^\s_{\a\m}\ ;\label{Riemann}\\
R=g^{\a\n}R^\m_{\ \a\m\n}\ .\qquad\qquad\qquad\eel{Ricci}
 In perturbation theory, one writes
 \be g_{\m\n}=\d_{\m\n}+\e h_{\m\n}\ ,\eel{gperturb}
where \(\e\) is taken to be the perturbation expansion parameter, and \(\d_{\m\n}\) is the Lorentz
invariant identity matrix. The inverse of this is
 \be g^{\m\n}=\d^{\m\n}-\e h_{\m\n}+\e^2 h_{\m\a}h_{\a\n}-\cdots\ , \eel{ginverse}
Coordinate reparametrization consists of the substitution of the coordinates \(x^\m\) as follows:
 \be x^\m\ra x^\m+\e\et^\m(x) ,\eel{xtransf}
where \(\et^\m(x)\) is the space-time dependent generator of this `gauge transformation'. The
metric tensor then transforms as
 \be g_{\m\n}\ra g_{\m\n}+\e\left(\et^\a\pa_\a g_{\m\n}+g_{\a\n}\pa_\m\et^\a+g_{\m\a}\pa_\n\et^\a\right)\ ,
\eel{gtransf} which for the \(h_{\m\n}\) fields implies
 \be h_{\m\n}\ra h_{\m\n}+D_\m\et_\n+D_\n\et_\m\ ,\eel{htransf} where we used the notion of a covariant
derivative, while indices are raised and lowered with the metric tensor \(g_{\m\n}\):
 \be \et_\m=g_{\m\n}\et^\n\ ;\qquad D_\m \et_\n\equiv \pa_\m \et_\n-\G^\a_{\m\n}\et_\a\ . \eel{covderiv}
 \def\lra{\leftrightarrow}

To expose the physical degrees of freedom describing gravitational radiation (gravitons), one
must first fix the gauge freedom. For our purposes here, the radiation gauge works fine:
 \be \sum_{i=1}^3\pa_i h_{i\m}=0\ ;\qquad\m=1,\cdots,4. \eel{radiation}
 Subsequently, we expand the action \eqn{EH} in powers of \(\e\). After removing pure derivatives
one finds that the terms linear in \(\e\) vanish. If we identify
 \be\e^2=16\pi G\ , \eel{Newtoncnst}
 the second order terms take the form
 \be\LL=-\half h_{\a\b}U_{\a\b\m\n}h_{\m\n} \ , \eel{secondorder}
where \(U_{\a\b\m\n}\) is a space-time operator containing second order derivatives. In momentum
space, this operator is found to take the form
 \be U_{\a\b\m\n}=\half k^2(\d_{\a\m}\d_{\b\n}-\d_{\a\b}\d_{\m\n})+ k_\m k_\n\d_{\a\b}-k_\b k_\n\d_{\a\m}+b^2\vec
k_\b\vec k_\n\d_{\a\m}\ ,\eel{Vrad}
 where \(\vec k\) is \(k\) with its time component replaced by 0, and the parameter \(b^2\) is sent
to infinity, so as to impose Eq.~\eqn{radiation}. To see the physical modes that propagate, in the
case that \(\e\) is infinitesimal\fn{When (classical) gravitational radiation is described, \(\e\)
indeed is exceptionally small, so that this perturbative description is extremely accurate. The
reason why higher order \(\e\) corrections are harmless is that the operator \(U\) has an inverse
(Eq.~\eqn{tensorprop}). The physics of these higher order \(\e\) corrections, which include the
effects of the energy transported by gravitational waves, is well understood.}, it is instructive
to rotate the spacelike components of the momentum vector into the positive \(z\) direction:
 \be \vec k_\m=(0,\ 0,\ \k,\ 0)\ . \eel{kz}
To find the propagator in this gauge, we first have to symmetrize \(U_{\a\b\m\n}\) with respect to
interchanges \(\a\lra\b\), \(\m\lra\n\) and \((\a\b)\lra(\m\n)\). The propagator \(\Bbb P\) is then
solved from
 \be {\Bbb U}\cdot{\Bbb P}={\Bbb I}\ ;\qquad {\Bbb I}=\half(\d_{\a\m}\d_{\b\n}+\d_{\a\n}\d_{\b\m})\ .
\eel{propeq}
 The solution to this tensor equation is
  \be P_{\m\n\a\b}(k)={1\over k^2-i\e}\left(\hat\d_{\a\m}\hat\d_{\b\n}+\hat\d_{\a\n}\hat\d_{\b\m}-{2\over
n-2}\hat\d_{\a\b}\hat\d_{\m\n}\right)+ \nm\\ \hbox{terms containing only}\ \vec k^2\ \hbox{in their
denominators,} \eel{tensorprop}
 where \(\hat\d\) is defined as
 \be\hat\d_{\m\n}\equiv\hbox{diag} (1,\ 1,\ 0,\ 0)\ ,\eel{deltahat}
and \(n\) is the number of space-time dimensions, \(n=4\) being the physical value. Only the part
explicitly written in Eq.~\eqn{tensorprop} represents excitations that actually propagate. One sees
first of all that only the completely transverse components of the field \(h_{\m\n}\) propagate:
\(\m,\n=1\hbox{ or }2\). Secondly, the diagonal component (the trace) drops out:
 \be P_{\m\m\,\a\b}=0\qquad\hbox{since}\quad \hat\d_{\m\m}=n-2\ . \eel{traceprop}
Henceforth, \(n=4\). Since then traceless, symmetric \(2\times2\) matrices have only two independent
components, we read off that there are only two propagating modes, the two helicities of the graviton. The
propagator \eqn{tensorprop} propagates a graviton with the speed of light.

There is one very important feature that we have to mention when treating perturbative gravity. The action is
not bounded when we make a Wick rotation to Euclidean space, in contrast with the boundedness of the action
for all other field theories in Euclidean space. This means that the bilinear terms in Eq.~\eqn{secondorder}
contain parts that have an unconventional sign. This unusual sign does not invalidate gravity as a field
theory, because the unphysical parts do not propagate. It does however have important implications: one is
that gravity has a potential instability; stationary gravitational fields carry a negative energy density, so
that gravitational collapse can occur. Another implication is that, in Euclidean space, one is not allowed to
demand the functional integral to go over the real values for all components of the metric \(g_{\m\n}\), even
after fixing the gauge. One will always encounter components of \(g_{\m\n}\) that will have to be integrated
over complex contours, so as to ensure convergence of the functional integrals.\fn{We are discussing the
functional integrals prior to the integrations over momentum space, so the divergence mentioned here is
distinct from the usual ultraviolet difficulties of quantum gravity.} The complex components of \(g_{\m\n}\)
are usually in the trace or dilaton components, and in the gauge fixing procedures they act as Lagrange
multipliers. Gravity generates Lagrange multiplier fields even in Euclidean space, where all functional
integrals would be real Gaussian integrals. In a functional integral, however, Lagrange multipliers must be
handled as imaginary fields in order to guarantee convergence. This can be shown to lead to the observation
that, even after Wick rotating to Euclidean space, one \emph{must} treat the dilaton sector of the
gravitational metric field as a complex field variable in the functional integration procedure, a fact
frequently overlooked in the literature. In our work, this feature causes a considerable complication, which
we shall handle in Section~\ref{indefinite.sec} and onwards.

Massless scalars are introduced in \eqn{EH} as
 \be \LL_1 =\LL_\f+\LL^\mathrm{matter}\ ,\quad \LL_\f=-\half\sqrt g\, g^{\m\n}\pa_\m\f^a\pa_\n\f^a\ ,
 \eel{masslessphi}
where \(a=1,\,\cdots,\,N\) is an internal index. \(\LL^\mathrm{matter}\) will contain all
contributions from other kinds of matter. Soon, we shall restrict ourselves to the case
\(N=n=4\), the dimensionality of space-time. The propagators are the inverse of this bilinear
action, or
 \be P^{ab}(k)={\d_{ab}\over k^2-i\e}\ . \eel{scalarprop}
It is imperative that the sign of all terms in the propagator be positive. This is because the residues of
the poles define the normalization of the one-particle states. If a residue is not one but some number \(Z\),
then the \(S\)-matrix elements generated by the Feynman rules will replace the ket-bra product \(|k\ket\bra
k|\) by \(|k\ket Z \bra k|\). One can renormalize these states by a factor \(\sqrt{Z}\), but one cannot
change the sign of \(Z\) this way. One can also note that \(Z\) is directly proportional to the probability
that the particle with momentum \(k\) is produced in some process. This probability must be a positive (or
vanishing) number. Having negative \(Z\) would necessitate the inclusion of indefinite norm states in the
quantum system. In classical field theories, fields with the wrong sign in the kinetic energy part of the
action would tend to destabilize the vacuum by allowing for negative energy states. Note that the
gravitational propagator, Eq.~\eqn{tensorprop} has only positive residues at its poles. This implies,
\textit{inter alia}, that conventional gravity as a field theory obeys unitarity, whereas it would have
generated indefinite norm gravitons if Newton's constant had been negative.

The masslessness of the scalars in Eq.~\eqn{masslessphi} is protected by a symmetry, the
Abelian Goldstone symmetry,
 \be \f^a\ra\f^a+C^a\ , \eel{Goldst}
where \(C^a\) is a set of constants. Indeed, these scalars are allowed to interact \emph{only}
gravitationally, otherwise they would have collected mass terms.

There is also a rotational symmetry, \(O(N)\), allowing us to rotate these scalars among one another. One
would not expect the scalars to have a Lorentz symmetry \(SO(3,1)\), but later we shall explain how this
symmetry nevertheless might come about.

\newsecl{Breaking the symmetry spontaneously}{break.sec}

For comparison, consider the case of a vectorial non-Abelian gauge theory,
 \be \LL=-\quart F_{\m\n}F_{\m\n}-\half(D_\m\f^aD_\m\f^a)-V(\f^2)\ , \eel{YML} where \(D_\m\) is the covariant
derivative containing the vector field \(A_\m\), and
 \be  F_{\m\n}=\pa_\m A_\n-\pa_\n A_\m+ \OO(gA^2)\ . \eel{covcurl}
Furthermore, the function \(V(\f^2)\) describes the self interaction of the Higgs field. In the
symmetric mode, gauge fixing could be done also by imposing the radiation gauge:
 \be \pa_i A_i=0\ , \eel{radgaugevect}
where we sum only over the spacelike components; as in Eq.~\eqn{radiation}, Latin indices only take
the values 1,2 or 3. The bilinear part of the action is then
 \be -\half A_\m U_{\m\n}A_\n\quad=\quad -\half(\pa_\m A_\n)^2+\half(\pa_4A_4)^2-\half b^2(\pa_iA_i)^2\
 , \eel{quadractvect}
where \(b\) tends to infinity so as to obey \eqn{radgaugevect}. In momentum space\fn{The factors
\(i^2\) are cancelled by the minus sign in the partial integration.},
 \be U_{\m\n}=k^2\d_{\m\n}-k_4^2\d_{\m4}\d_{\n4}+b^2\vec k_\m\vec k_\n\ . \eel{Uvec}
Choosing \(\vec k\) to be in the \(z\)-direction, Eq.~\eqn{kz}, this matrix is
 \be U_{\m\n}=\hbox{diag}(k^2,\,k^2,\,k^2+b^2,\,\vec k^2)\ . \eel{Uvecdiag}
In the limit \(b\ra\infty\), its inverse is the vector propagator in the radiation gauge,
 \be P_{\m\n}(k)={\hbox{diag}(1,1,0,0)\over k^2-i\e} + \hbox{the Coulomb part containing only }1/\vec k^2\ .
 \eel{vecprop}
Thus, we see that only two helicities for the photon are propagated along the light cone. In
addition, we have the scalars with the usual propagators.

The BEH mechanism is now looked at in the following way. Assuming that the potential \(V\) forces the scalar
field \(\f\) to take a non-vanishing vacuum value,\fn{Strictly speaking, this mechanism has little to do with
the vacuum; the value of the scalar field is nearly never exactly vanishing, so it could always be used to
fix the gauge. However, if we arrange things such that this value is already unequal to zero at lowest order
in perturbation theory, then we see how this mechanism works order by order in the perturbation expansion,
which makes it much more transparent. Indeed, if one attempts to describe a non-perturbative version of the
BEH mechanism several complications arise that we do not wish to go into.} it can be used to define a
preference gauge. The preference gauge is the gauge in which \(\f\) is pointing in a constant direction, so
that its vacuum value, \(\bra\f\ket\), is constant. If in this gauge \(A_\m=0\), then we have \(D_\m\f=0\).
Consequently, the kinetic term of the \(\f\) field, \(-\half D_\m\f^2\), acts as a mass term for the gauge
field in this gauge. Indeed, if we use this scalar to fix the gauge, thus replacing the radiation gauge, the
gauge field propagator becomes
 \be P_{\m\n}(k)={\hbox{diag}(1,1,1,0)\over k^2+M^2-i\e}\quad \hbox{in the Lorentz frame where }
 k_\m=(0,0,0,k)\ ,  \eel{massvecprop}
where \(M\propto g\bra\f\ket\). We observe that each vector particle (we may have one or more of
them), obtains a mass term, borrowing its third degree of freedom from the scalars. The scalars
lose one degree of freedom for each vector particle that gets a mass, because it is used to fix the
gauge.

Now let us do the same thing for the gravitational theory. For simplicity, we restrict the
discussion to the case of a four-dimensional spacetime. Four scalar fields \(\f^a,\
a=1,\dots,4\), will be used to fix the gauge. Fixing the gauge means fixing a preferred
coordinate frame. The only way to do this is, to identify these four coordinates. Therefore,
we write
 \be \f^\m=m\,x^\m\ ,\qquad\m=1,2,3,4\,,\eel{coordfix} where the parameter \(m\) is yet to be
determined. This immediately raises a question: the four coordinates are in Minkowski space; if all
four scalars are real fields, the kinetic term \eqn{masslessphi} leaves a Euclidean \(O(4)\)
symmetry, not the Lorentz group \(O(3,1)\). The residual system, after spontaneous symmetry
breakdown, should still exhibit Lorentz invariance, not the Euclidean rotation group.

The answer to this question is postponed for a moment. Here, we simply assume that \(\f^4\) is
shifted by an \emph{imaginary} shift \(i\, m t\). For the time being, let us note that the
Goldstone symmetry, \eqn{Goldst}, actually allows the constants \(C^a\) to be complex.
Shifting the fields \(\f^a\) by complex rather than real numbers, also leaves the action
invariant (after partial integrations). We may also note that the quantum mechanical
functional integral, \(\int\DD\f^a\), allows a shift of the integration contour into the
complex plane:
 \be \int\DD\f^a=\int\DD(\f^a+C^a)\ , \eel{contourshift}
which means that, in the action, we are allowed to shift the fields by any complex parameter
\(C^a\).\par \emph{Shifts of field variables by a complex parameter are allowed if the action
itself stays real!}\par\noindent This is the case if the action is even in these field variables,
as it indeed is here. Consequently, we are encouraged to investigate the gauge choice
 \be \f^1=m\,x\,,\ \f^2=m\,y\,,\ \f^3=m\,z\,,\ \f^4=i\,m\,t\ . \eel{complexgauge}
This last condition stems from the contour shift \(\f^4=imt+\d\f^4\), after which the real field
\(\d\f^4\) was gaugefixed to be zero.

There may, however, be a more practical solution to this problem that we can mention here. In
perturbation theory, we find that, if instead of \(\f^4\) we take \(\f^0\) to be real, the original
Lagrangian \eqn{masslessphi} can still be used. The intermediate states in the unitarity
condition\cite{DIAGRAMMAR} for the \(S\)matrix receive a minus sign. These signs however are
multiplicative. There is a plus sign if we restrict ourselves to intermediate states with an even
number of \(\f^0\) particles. Fortunately, the Lagrangian \eqn{masslessphi} is even in \(\f^0\).
Therefore, the odd states do not occur! We shall refer to this as the `pairing mechanism'.

Again, the question is, is this unorthodox procedure legal? In particular, are we allowed to have a
time-dependent vacuum expectation value, which also breaks the \(\f^0\leftrightarrow -\f^0\)
symmetry? We shall check the law as we go along. The penalty for this unorthodox procedure will
turn out to be considerable, but we will see what the consequences are and what actually can be
done to repair the theory later.

There is another problem with the gauge choice \eqn{complexgauge}: it is not free from ambiguities.
What if the scalar fields take the same values at different points in space-time? Curiously, this
problem will turn out to be related to the previous one. For the time being, let us note that there
is a natural way to restrict ourselves to the unambiguous sector of the theory by imposing the
condition that the volume element
 \be \e^{\m\n\a\b}\pa_\m\f^1\pa_\n\f^2\pa_\a\f^3\pa_\b\f^4 \ >\ 0\ . \eel{volpos}
We will see how this condition emerges in the cure for the indefinite metric problem.

\newsecl{Spontaneously broken gravity}{massive.sec}

The gauge constraint \eqn{complexgauge} completely removes the four scalars as physical degrees of freedom.
We now ask what the new action will look like in this gauge. We expect that general covariance will appear to
be absent, and that the new gravitational field will have 2\,+\,4\,=\,6 physically propagating modes.

The contribution from \(\LL_1 \), Eq.~\eqn{masslessphi}, is derived as follows:
 \be \pa_\m \f^\n=m\pa_\m x^\n =m\d_\m^\n \ , \crl{Dmuphi}
 \LL_\f =-\half m^2\sqrt{g}\ g^{\m\m}\ . \eel{massterm}
Here, because of the \(i\) in the gauge choice \eqn{complexgauge}, the summation over the Lorentz
indices is the Lorentz covariant one (for simplicity, the Euclidean notation is used throughout).
Of course we notice the first violations of general covariance. Writing
 \be g_{\m\n}=\d_{\m\n}+\e h_{\m\n}\ ;&& g^{\m\m}=4-\e h_{\m\m}+\e^2(h_{\m\a})^2+\OO(\e h)^3\ ;
 \quad\sqrt{g}\ = \nn
 \exp({\half\mathrm{Tr}\log(\d_{\m\n}+\e h_{\m\n})})\!\!&&=\ 1+\half\e h_{\m\m}-\quart\e^2
 (h_{\m\a})^2\hat{}+\fract18\e^2(h_{\m\m})^2+\OO(\e h^3)\ ,\quad{} \eel{determnt}
we find that
 \be \LL_1 =-2m^2-\half \e m^2h_{\m\m}+0+\OO(\e h)^3  \eel{Lmatterh}
It so happens that, in Eq.~\eqn{Lmatterh}, the terms of second order in \(\e\) cancel out.
There is however a part linear in \(h_{\m\n}\), which would give rise to `tadpole diagrams' in
the quantum theory, and inhomogeneous equations for \(h_{\m\n}\) in the classical perturbation
expansion. These would cause shifts in the \(h_{\m\n}\) fields, which is inadmissible because
we had already subtracted the `vacuum part' of the metric field \(g_{\m\n}\). What this means
is that the non-vanishing expectation values \eqn{coordfix} would be eliminated by the
dynamics of the theory.

However, there is another way to remove the term linear in \(h_{\m\n}\). Let us now add a
cosmological constant \(\L\) to the action in the symmetric representation. Thus, we have a
third term in the total Lagrangian:

\be \LL^\L\iss{-\L\over 8\pi G}\sqrt{g}\iss{-\L\over 8\pi G}(1+\half\e h_{\m\m}-\quart\e^2
 (h_{\m\a})^2\hat{}+\fract18\e^2(h_{\m\m})^2+\OO(\e h^3)\ . \eel{Lcosmo}
So we can absorb the linear term by having a (negative) cosmological constant,
 \be \L=-8\pi G m^2\ , \eel{ccnst}
in the unbroken Einstein-Hilbert action \eqn{EH}. What we see is that it cancels the
energy-momentum tensor of the vacuum values of the scalar fields.

We now get the picture. The starting point is an Einstein-Hilbert action with scalar fields and a
negative cosmological constant. We search for a solution that has a vacuum state, which means that
the universe obeys asymptotically flat boundary conditions at \(\infty\). Such solutions do exist
but require the scalar fields to vary with space and time, because they must generate an
energy-momentum tensor to neutralize the effects of the cosmological constant. These space-time
dependent vacuum values of the scalar field spontaneously break reparametrization invariance. Now,
let us see what this does to the graviton.

The surviving action, including \eqn{Lmatterh} and \eqn{Lcosmo}, is
 \be \LL={\sqrt{g}\,R\over 16\pi G}-m^2+\e^2m^2(-\quart h_{\m\a}^2+\fract18 h_{\m\m}^2)+\OO(\e h^3)\ . \eel{Ltot1}
Let us expand the first term as well, remembering now that no further gauge fixing should be
allowed,
 \be {\sqrt{g}\,R\over 16\pi G}=\e^2\Big(-\quart(\pa_\m h_{\a\b})^2+\fract18(\pa_\m h_{\a\a})^2+\half A_\m^2\Big)
 +\OO(\e h)^3\ , \eel{EHexp} where \be A_\m=\pa_\a h_{\a\m}-\half\pa_\m h_{\a\a}\ . \eel{Aterm}
We go again to Fourier space. We can read off from Eq.~\eqn{Ltot1}, which does not vanish for any
of the components of \(h_{\m\n}\) at zero momentum, that there will be no singularity at vanishing
\(k^2\). It therefore makes sense to rotate the 4-vector \(k\) to the Euclidean 4 direction:
\(k_\m=(0,0,0,k)\). We ignore the \(-m^2\) in \eqn{Ltot1}, and identify the remainder as a mass
term.

The combined Lagrangian, obtained by substituting \eqn{EHexp} in \eqn{Ltot1} in momentum space, can
now be written as
 \be \LL=(k^2+m^2)(-\quart h_{\a\b}^2+\fract18 h_{\a\a}^2)+\half k^2(h_{4\m}-\half h_{\a\a}\d_{4\m})^2\ ,
 \eel{Ltot2}
where we ignored the higher order terms and the \(-m^2\). Now we split the dynamical variables into
irreducible parts. As before, Latin indices \(i,\,j,\,k,\,\dots\) run from 1 to 3.
 \be &&h_{ii}=u\ ,\qquad h_{ij}=\tl h_{ij}+\fract13 u\,\d_{ij}\ ,\qquad \tl h_{ii}=0\ , \nn
 &&h_{i4}=h_{4i}=h_i\ ,\qquad  h_{44}=h\ . \eel{compons}
In terms of the new variables \(\tl h_{ij},\ h_i,\ h\) and \(u\), we find
 \be \LL=(k^2+m^2)(-\quart\tl h_{ij}^2+\fract16 u^2)-\half m^2 h_i^2-\fract18 m^2(h-u)^2\ . \eel{massiveL}
The fields \(\tl h_{ij}\) are the 5 components of a massive spin 2 particle. The field \(u\) is
clearly also a dynamical field, whereas \(h_i\) and \(h-u\) do not propagate, since they have no
pole in the propagator.

\newsecl{Massive gravity, the general case}{massive2}

The Lagrangian \eqn{Ltot1} is not the most general Lagrangian for massive spin 2 fields. One
can ask instead, what is the most general bilinear Lagrangian for symmetric fields
\(h_{\m\n}\) such that inverting the bilinear coefficients leads to a propagator that has
single poles in the momentum \(k\), corresponding to a spin 2 particle with \(m\), and a spin
0 particle with mass \(\m\)? This is a fairly complicated exercise, but the result of the
calculation is quite elegant and simple to explain.

Let us assume that these poles occur exclusively in those components of the propagator that are orthogonal to
the momentum \(k_\m\). This implies that, in the residue of the pole term of the propagator, all indices of
the \(h_{\m\n}\) fields must be contracted by the ``spacelike Kronecker delta" \(\d_{\m\n}+k_\m k_\n/m^2\),
and consequently, in the limit \(m^2\ra0\), the propagator diverges inversely with \(m^2\) as soon as one of
the indices of the \(h_{\m\n}\) field is parallel to \(k_\m\) or \(k_\n\). This implies that, in the limit
\(m^2\ra 0\), the bilinear part of the Lagrangian is invariant under the replacement \be h_{\m\n}\ra
h_{\m\n}+\pa_\m \eta_\n+\pa_\n\eta_\m\ , \eel{newgaugetrf} where \(\eta\) is infinitesimal. Of course, this
is nothing but the gauge transformation \eqn{htransf} in the perturbative regime. One concludes that the part
of the Lagrangian that contains two derivatives coincides exactly with the bilinear part of the
Einstein-Hilbert action, Eq.~\eqn{EHexp}. This is indeed what one gets in an explicit calculation.

For the mass terms, there are two possibilities left, so that one may choose two coefficients,
\be\LL^m=-\quart m_1^2 h_{\m\n}^2+\quart m_2^2 (h_{\a\a})^2\ . \eel{massterms} In the case \(m_1=m\,,\
m_2^2=\half m^2\), this is the Lagrangian \eqn{Ltot2} obtained from spontaneous symmetry breaking. Note that
we anticipate a scalar with an overall wrong sign in its propagator. What are the masses \(m\) and \(\m\) in
terms of \(m_1\) and \(m_2\) in the general case?

To find out, we consider the Lagrangian in momentum space and rotate the momentum into the
time direction, \(k_\m=(0,0,0,k)\). Again let Roman indices \(i,j,\dots\) take the values 1,2,
and 3 only. Decomposing \(h_{\m\n}\) again as in Eqs.~\eqn{compons}, we find
 \be\LL&=&-\quart(k^2+m_1^2)(\tl h_{ij})^2-\half m_1^2\, h_i^2+{1\over
 4(m_2^2-m_1^2)}\Big((m_2^2-m_1^2)h+m_2^2\,u\Big)^2+\nn &&\fract16(k^2+\m^2)u^2\
 ,\qquad \hbox{with}\qquad {2\m^2\over m_1^2}={4m_2^2-m_1^2\over m_1^2-m_2^2}\ . \eel{mumone}
In this expression the various terms were rearranged such that we can easily read off the masses. The fields \(h_i\)
and \(h\) have no kinetic terms, and so they do not propagate with the pole of a physical particle. In the absence of
source terms or higher order terms, the field equations enforce \be h_i=0\ ,\qquad h={m_2^2\over m_1^2-m_2^2}\,u\ .
\eel{hih} The particle \(u\) does propagate, with the wrong sign in its propagator, and mass \(\m\) given by
Eq.~\eqn{mumone}. The spin 2 fields have mass \(m=m_1\). We have \(\m=m=m_1\) in the case of spontaneous symmetry
breaking, where \(m_2^2=\half m_1^2\) and Eq.~\eqn{mumone} coincides with \eqn{massiveL}.

It is important now to note that, if \(m_1=m_2\), the \(u\) field gets an infinite mass. This
indeed is the Fierz-Pauli case\cite{FP}\cite{vDV_1970} Only in this case, the effects of the
unphysical \(u\) particle disappear, and we have an explicitly unitary theory. It would be
ideal if we could generate this Lagrangian from a BEH mechanism, but, within our formalism,
this is unlikely. Unfortunately, we could not find a BEH mechanism of this sort. It appears
that the above unitary Lagrangian cannot emerge because the matter fields required would have
to consist of real fields that generate an energy-momentum in the spontaneously broken phase
that is proportional to \(g_{\m\n}\). To achieve this, unconventional matter fields would be
required; scalar fields --- or vector fields --- will not do.

\newsecl{Removing indefinite metric states}{indefinite.sec}

We return to Eq.~\eqn{massiveL}. The field \(u\) stands for a scalar particle, the
\(6^\mathrm{th}\) degree of freedom, as expected, and it has the same mass as the heavy
graviton. But, of course, the reader will see what the problem is. The field \(u\) has the
wrong sign in its propagator:
 \be P^{uu}(k)={-3\over k^2+m^2-i\e}\ . \eel{uprop}
Consequently, the theory we have arrived at now, will violate unitarity. The wrong sign here is
directly related to the indefiniteness of the action in Euclidean space that we referred to in
Section~\ref{EHgrav.sec}, and further analysis shows that it is also related to the fact that our
fourth scalar field has the wrong sign in its kinetic term if used as a timelike component.

In the symmetric representation, however, the vacuum appears to be stable, due to the pairing
mechanism, and one might wonder whence the instability of the vacuum after the shift \(\f^\m\ra
mx^\m+\tl \f^\m\). Since the role of the \(\f^0\) field is taken over by the \(u\) field, one might
suspect that the \(u\) field should be taken to be imaginary, not real. Only then we see that two
massive particles emerge, one with spin 2 and one with spin 0.

To study a possible pairing mechanism for the \(u\) field, let us now concentrate on the other
terms in the Lagrangian. The matter fields couple to the gravitational fields through the energy
momentum tensor \(T^{\m\n}\) of the matter field. We write \(\LL^\mathrm{matter}(g_{\m\n})\) to
indicate the dependence of this contribution on the metric \(g_{\m\n}\):
 \be  g_{\m\n}=\d_{\m\n}+\e h_{\m\n}\ , \qquad \LL^\mathrm{matter}(g_{\m\n})=\LL^\mathrm{matter}
 (\d_{\m\n})+\half\e T^{\m\n}h_{\m\n}+\OO(\e^2)\ . \eel{matterterm}
In momentum space, after again rotating the momentum vector into the 4 direction,
\(k_\m=(0,0,0,k)\), we write (ignoring higher order corrections),
 \be T=T^{\m\m}\ ,&& T^{\m\n}=T_\bot^{\m\n}+\quart T\d^{\m\n}\ ,\qquad T_\bot^{\m\m}=0\ ,\crl{TraceT}
 \pa_\m T^{\m\n}=0&\ra&T^{4\m}=0\ ,\qquad T_\bot^{44}=-\quart T\ , \qquad T^{ii}= T\ ,\\
 T^{ij}&=&\tl T^{ij}+\fract13 T \d^{ij}\iz\tl T^{ij}_\bot+\quart T\d^{ij}\ ,\qquad \tl T^{ii}=0\ .
 \eel{energymomentum}
Thus, we find that the coupling term is
 \be \half T^{\m\n}h_{\m\n}\iss\half\ T_\bot^{ij}+\fract18 T u\iss\half\ \tl T^{ij}\tl h_{ij}+
 \fract16 T u\ . \eel{Tucoupling}

So, the \(u\) field is coupled to the trace of the energy-momentum tensor. It
is this coupling that causes violation of unitarity. Why do we have violation
of unitarity? In the symmetric representation, we could see that the pairing
mechanism filters out the negative metric states. Unfortunately, that mechanism
now fails. We suspect that the reason for this failure is that the gauge
choice, Eq.~\eqn{complexgauge}, also selects out a vacuum state that breaks the
\(\f^0\leftrightarrow -\f^0\) symmetry. Therefore, the vacuum mixes positive
and negative metric states; thus, we cannot select out the even states, as was
our intention. The pairing mechanism for the \(\f^0\) field fails. One may also
observe that we chose a vacuum value for \(\f^4\) that is time-dependent, and
this may imply a breakdown of energy conservation, or, indeed, a new
instability of the vacuum.

Imposing now the pairing mechanism for the \(u\) field implies a new constraint on our theory: the
\(u\) field should decouple. It was proposed in Refs.~\cite{Kaku}\cite{Porrati}, where similar
theories are discussed, that we should limit ourselves to having only ``conformal matter", that is,
matter with vanishing trace of the energy momentum tensor, \(T=0\). However, in cosmology this
would be a somewhat mysterious restriction, whereas in QCD we definitely do not want to consider
only scale invariant states. We conclude that our theory requires a modification.

As stated earlier, conventional matter such as scalar, spinor or vector fields (without analytic continuations to
complex field values) can never do the work right. This is because we wish to have a flat spacetime after symmetry
breakdown, and only a cosmological constant to our disposal to cancel the effects of the energy-momentum distribution
\(T^{\m\n}_0\) of our background fields, but ordinary matter never gives a \(T^{\m\n}\propto g^{\m\n}\) as in dark
energy.

One alley that we investigated is whether we can add a dilaton field \(\eta(x)\) to our
theory. The Lagrangian is then replaced by
 \be  &\LL\ra {\sqrt g\over 16\pi G}\left(Re^{\a\eta}-2\L e^{\b\eta}\right)+\LL_\f
 e^{\g\eta}+\LL^\mathrm{matter}e^{\k\eta} -\half\sqrt g\, e^{\l\eta}g^{\m\n}(\pa_\m\eta\pa_\n\eta)
 -\half\sqrt g\, m_3^2\eta^2\ ,& \nn \eel{dilaton}
where \(\a,\,\b,\,\g,\,\k,\,\l\), and \(m_3\) are constants that can be adjusted. It turns
out, however, that there are no possible choices for these constants such that either the
coupling to the \(u\) field vanishes or that the kinetic terms all get the correct sign, nor
could the mass of the \(u\)-field be sent to infinity. This alley was a blind one.

Next, we investigate whether an other observation might lead to solutions to this problem: the scalar fields
\(\f^a\) themselves actually generate an `alternative' metric tensor:
 \be g^\f_{\m\n}=\pa_\m\f^{\,a}\pa_\n\f^{\,a}\ . \eel{gphi}
This is the metric tensor of flat space, in the gauge \eqn{complexgauge}. It is a new tensor obeying the same
transformation rules as \(g_{\m\n}\). Matter could be coupled either to \(g_{\m\n}\) or to \(g^\f_{\m\n}\),
and there is no a priori reason to select any particular choice. In a renormalizable theory, there would be a
reason: the derivatives in Eq.~\eqn{gphi} would render couplings with \(g^\f_{\m\n}\) highly
non-renormalizable. But, we had to give up renormalizability from the very start, as we are dealing with
gravity.

One attempt to resolve the metric problem, based on this idea, is as follows. Impose a constraint on the
scalar fields \(\phi\):
 \be g^\f=g\ , \eel{gconstr}
which is invariant under coordinate transformations, and can also be written as
 \be \e^{\m\n\a\b}\pa_\m\f^1\pa_\n\f^2\pa_\a\f^3\pa_\b\f^4=\sqrt g\ . \eel{volconstr}
In the gauge \eqn{complexgauge}, we now have \(g=1\), hence \(h_{\a\a}=u+h=0\).

Unfortunately, upon closer inspection, this procedure does not do the job right. Substituting \(u=-h\) in
Eq.~\eqn{massiveL}, we find that now the \(h\) field has become a dynamical field, with the wrong sign in the
kinetic term of its propagator. Since the gauge choice \eqn{complexgauge} was already made, and supposed to
be unitary, the \(h\) field is not a ghost here, so it produces physical states with the wrong metric. \(h\)
is coupled to \(T^{44}\), which does not vanish.

Alternatively, one could consider the following approach. Let us couple matter not with the metric
\(g_{\m\n}\) of our gravity sector, but with
 \be g^\mathrm{matter}_{\m\n}= g^{-\fractje14}g_{\m\n}\ (g^\f)^\fractje14\ , \eel{gmatter}
where \(g^\f\) is the determinant of the metric \(g_{\m\n}^\f\). This means that the trace of
\(h_{\m\n}\) does not enter in the matter Lagrangian, but is replaced by the trace of the
perturbation of \(g^\f_{\m\n}\), and, since in our scalar-field gauge \eqn{complexgauge} this
vanishes, matter will not couple to the trace of \(h_{\m\n}\). Conversely then, one expects the
trace \(T\) of the energy-momentum tensor not to couple to the \(h_{\m\n}\) field.

This, however, is not correct.\fn{I thank Luca Vecchi for detecting this error in version \# 3 of
this paper.} The deeper reason why one has to reexamine the equations is that matter modified by
the insertion \eqn{gmatter} obeys conservation laws that differ from
Eqs.~\eqn{TraceT}--\eqn{energymomentum}.

Since \(h_{\a\a}=h+u\), the matter coupling \eqn{Tucoupling} will be replaced by
 \be \half T^{\m\n}(h_{\m\n}-\quart h_{\a\a}\d_{\m\n})=\half T^{ij}_\perp \tl h_{ij}+
 \fract1{24}u T-\fract18 h T\ . \eel{Thcoupling}
Indeed, the coupling with \(u\) is replaced by a coupling with the combination \(\fract13 u-h\), but, since the
dynamical equations from the Lagrangian \eqn{massiveL} force \(h=u\) (apart from a contact term), one finds that the
\(u\) field couples with matter after all. Let us now, however, replace Eq.~\eqn{gmatter} by
 \be g^\mathrm{matter}_{\m\n}= g_{\m\n}\ (g^\f/g)^\a\ , \eel{gmatterprime} where the coefficient
 \(\a\) is yet to be determined. We then we get the coupling
 \be \half T^{\m\n}(h_{\m\n}-\a h_{\a\a}\d_{\m\n})=\half T^{ij}_\perp \tl h_{ij}+((\fract16-\half\a)u-\half\a
 h) T\ , \eel{Tuminhcoupling} and this time we can ensure that only the contact term survives. Since
the propagator generated by the kinetic term \eqn{massiveL} enforces \(h=u\) (apart from contact
terms), the condition for this to happen is that \be \a=\fract16\ . \eel{alphafixed}

Thus, we can in principle avoid the direct couplings of matter with single \(u\) particles so that, at least
in the direct channel, unitarity is restored. As long as the \(u\) particles are only pair created, they do
not violate unitarity. Unfortunately, we cannot exclude odd terms in the \(u\) field at higher orders in the
pure gravity sector, a difficulty that needs to be investigated further.

Note, that these proposals require \(g^\f\) not to vanish. If we write
 \be \f^\m=mx^\m+\tl\f^\m\ , \eel{phipert}
a perturbative expansion gives
 \be \sqrt{g^\f}=m^4+m^3\pa_\m\f^\m+\half m^2((\pa_\m\f^\m)^2-\pa_\m\f^\n\pa_\n\f^\m)+\OO(\f^3)\ ,
\eel{gphiexp} which can be derived elegantly,
 \be \sqrt{g^\f}=\mathrm{det}(\pa_\m\f^a)=\e^{\m\n\a\b}\pa_\m\f^1\pa_\n\f^2\pa_\a\f^3\pa_\b\f^4\ ,
 \eel{volumedet}
and this also shows that the expansion terminates beyond the \(m^0\f^4\) term. Note that this is
the volume term \eqn{volpos}, which now plays a more prominent role.

\newsecl{Relation to earlier work}{earlier}

Van Dam and Veltman\cite{vDV_1970} noted in 1970 that there are fundamental differences
between pure gravity, where gravitons have only two polarizations, and massive spin 2
theories, where the particle has 5 polarizations; their propagators are different already at
the tree level. They considered the unitarity requirement for the propagator, but made no
attempt to find the Lagrangian of a field theory that would generate such a propagator. The
discontinuity in the amplitudes that they found, is due to the fact that the massless theory
requires a gauge constraint, which is not the same constraint as what the theory is driven to
when mass terms are added; in fact, the unitary massive theory (with \(\m\ra\infty\) in
Eq.~\eqn{mumone}), enforces \(u=0\), or \(\pa_\m\pa_\n h_{\m\n}=\pa^2 h_{\m\m}\)
(perturbatively), which cannot be used as a gauge constraint because it is perturbatively
gauge-invariant (it is the condition \(R=0\) at lowest order).

After completion of the first version of this paper, various responses were received notifying us of other early work
and many further references. Omero and Percacci\cite{percacci} discussed a Higgs phenomenon in quantum gravity, using
the Palatini formalism. Their aim is to apply this mechanism to compactify extra dimensions, where the difficulty with
the timelike component does not occur. Percacci observes, like we do, that we have not yet achieved a Higgs mechanism
that is completely smooth in the UV direction; it is rather like that of a non-linear sigma model.

Of particular interest is the work by Kiritsis et al\cite{Kiritsis}, who are making progress in the AdS/CFT approach.
They report that, even though the gluonic sector of QCD is purely bosonic, one nevertheless may consider a superstring
here, with extra projection operators excluding the fermionic modes.

Arkani-Hamed et al\cite{arkani02} consider several tensor fields and multiple sets of general
coordinate transformations. Like Siegel\cite{Siegel93}, they view the graviton as a bound state in
open string field theory, but they too make no attempt at rigorously discussing unitarity, which
would drastically reduce the number of free parameters and the amount of non-locality that seems to
characterize these theories.

Chamseddine\cite{cham03} also considers spontaneous symmetry breaking, but obtains a massive
graviton interacting with a massless one. He also does not discuss unitarity in the longitudinal
sector.

The BEH mechanism for gravity has been speculated on also in\cite{Duff}---\cite{Leclerc}, usually in
connection with brane theories and/or cosmology; in our work we focussed on the problems associated with the
indefinite metric. A bridge between our observations and the AdS/CFT approach is further discussed in Bandos
et al\cite{Bandos}, who observe that a gravitational Higgs effect takes place on \(p\)-branes imbedded in
higher dimensions; gravity in the bulk is unbroken, but on the \(p\)-brane it is massive.

This modified paper is a considerable improvement of its earlier version. A mistaken idea was withdrawn, as
explained in Section~\ref{indefinite.sec}, and we added Section~\ref{massive2} to explain how a Lagrangian
for a unitary description of only spin 2 particles looks, which however could not be obtained within our
present scheme. In Section~\ref{earlier}, further references are briefly described. The author is indebted,
among others, to R.~Jackiw, M.~Duff, I.~Bandos, W.~Siegel, A.~Chamseddine, E.~Kiritsis and L.~Vecchi for
their comments.

\newsecl{Conclusion}{concl}

The equivalent of the Brout-Englert-Higgs mechanism for gravity may exist. In the symmetric representation,
four scalar fields are added to the gravitational degrees of freedom, and a negative cosmological constant is
added. After assuming space-time dependent vacuum values for the scalar fields, they rearrange to produce a
field theory for a massive spin 2 particle and a scalar. The scalar would have unphysical metric, so that it
has to be removed from the system. This can be done by modifying the matter part of the Lagrangian, so that
the zero spin field decouples. The ghost poles will cancel in the usual way by employing the Faddeev Popov
ghost, and using BRST imvariance. We do note that the insertion of Eq.~\eqn{gmatterprime} in the Lagrangian
for the matter field introduces higher derivatives there. This is only allowed in perturbation expansion,
which however will become more divergent in the ultraviolet. The theory was already non-renormalizable, so
this implies once again that we must constrain ourselves to some finite order in the perturbation expansion.

The complications in the longitudinal (scalar) sector always require a special treatment of
the volume factor in the metric, normally controlled by \(\sqrt g\). It is important to
realize that this can be done, at least at the level of classical, effective field theory. Our
models are quite singular in the ultraviolet region, but not yet all possibilities at getting
less divergent versions have been explored. Ideally, one would like to have a scenario where
the scalars only play their special role in the infrared domain, where the effects of the spin
2 mass are important, while they interact only mildly in the far ultraviolet. This, we hope,
might be something that could be realized in string theories. Indeed, recent AdS/CFT
approaches are pointing in this direction.

Imposing the pairing mechanism to the \(u\) field implies, in a sense, that the conformal sector of
gravity theory, described by the determinant \(g\) of the metric, must be treated in such a way
that this determinant is path-integrated in the complex direction: \be g_{\m\n}&=&\w\tilde
g_{\m\n}\ , \nn \det(\tilde g_{\m\n})&=&1\ , \nn \w=g^{1/4}&=& e^{i\eta}\ , \nn \tilde
g_{\m\n}=\hbox{ real}\ , && \eta=\hbox{ real}. \eel{detgcompl}

Our research is not complete. There are other potential difficulties. So-far, we only considered the coupling of the
\(u\) field to matter, and concluded that the matter Lagrangian had to be coupled to the conformal part of the metric
in an anomalous fashion --- the replacement of \eqn{gmatter} by \eqn{gmatterprime}, with \(\a=\fract16\), was an
important enough correction to warrant the submission of the 4th corrected version of this manuscript. But how can any
of these proposals be reconciled with the fact that the gravitons themselves are not conformally invariant? Newton's
constant has a non-trivial dimension. This difficulty is reflected in the fact that, although the \(u\) field does not
couple to matter, it does couple to itself, and we note that odd powers of \(u\) might arise in the gravitational self
couplings. Can these be renormalized away, as suggested above? The following consideration casts further doubts on the
validity of this assumption.

We can isolate the troublesome indefinite metric component of gravity by splitting the metric
tensor \(g_{\m\n}\) as in Eqs~\eqn{detgcompl} but without the \(i\):
 \be g_{\m\n}=e^{\et(x)}\tl g_{\m\n}(x)\ ,\qquad \et\ =\ \hbox{ real}; \nn \sqrt{\tl g}=1\ ,
 \qquad\sqrt g=e^{2\et}\ . \eel{detgreal}
The Einstein-Hilbert Lagrangian then becomes
 \be \LL^{\mathrm{EH}}\ =\ e^\eta(\tilde R+\fract32 \tilde g^{\m\n}\pa_{\,\m} \eta\pa_{\,\n}\eta)\ .
\eel{L-A} Apart from a normalization \(1/\sqrt{3}\), we could identify this \(\et\) field with the
field \(\f^0\), needed for the spontaneous breaking in the time direction, and start from the
Lagrangian
 \be \LL=e^\et\Big(\LL (\tilde g_{\m\n})-\half \tilde g^{\m\n}\left(\pa_{\,\m}\f^i\pa_{\,\n}\f^i -
 \pa_{\,\m}\f^0\pa_{\,\n}\f^0\right)\Big)\ , \eel{goodL}
where \(i\) counts the three spacelike components. This definitely has the right metric. However,
if we impose, as we would like to do,
 \be \bra\f^\m\ket\rightarrow m\,x^\m\ ,\qquad \eta\rightarrow m\,t/\sqrt{3}\ , \eel{corrvacexp}
we see that the prefactor for the Einstein term explodes exponentially with time: Newton's constant becomes
exponentially time dependent. This is not a small effect; the time scale in the exponent is of the order \(1/m\). This
is not flat space-time.

\end{document}